\documentclass[twocolumn,aps,prb,showpacs,reprint,showkeys]{revtex4-1}

\usepackage{times}

\usepackage{amsmath,amssymb}           

\usepackage{graphicx}                   
\usepackage{subfigure}                  

\DeclareTextSymbol{\degre}{OT1}{23}

\begin{document}

\title{Why the wire medium with finite dimensions makes a resonant metalens}
\author{Fabrice Lemoult}
\email{fabrice.lemoult@espci.fr}
\author{Geoffroy Lerosey}
\author{Mathias Fink}
\affiliation{Institut Langevin, ESPCI ParisTech \& CNRS, 10 rue Vauquelin, 75231 Paris Cedex 05, France}
\date{\today}

\begin{abstract}
This article is the first one in a series of two dealing with the concept of "resonant metalens" we recently introduced  [Phys. Rev. Lett. 104, 203901  (2010)]. Here, we focus on the physics of a medium with finite dimensions consisting on a square lattice of parallel conducting wires arranged on a sub-wavelength scale. This medium supports electromagnetic fields that vary much faster than the operating wavelength. We show that such modes are dispersive due to the finiteness of the medium. Their dispersion relation is established in a simple way, a link with designer plasmons is made, and the canalization phenomenon is reinterpreted at the light of our model. We explain how to take advantage of this dispersion in order to code sub-wavelength wave fields in time.
Finally, we show that the resonant nature of the medium ensures an efficient coupling of these modes with free space propagating waves and, thanks to the Purcell effect, with a source placed in the near field of the medium.    
\end{abstract}

\maketitle

\section{Introduction}

In a recent paper \cite{metalens}, we have introduced the concept of "resonant metalens" which can be used to break the diffraction barrier. We defined it as a cluster of coupled resonators, arranged on a subwavelength scale. When illuminated with broadband electromagnetic waves this macro-resonator radiates waves  that contain information of an object placed in its near field. By measuring this radiation one can recover sub-wavelength details of the object, and thus focus (or image) waves on a deep subwavelength scale, not limited by the Rayleigh criterion. 

In the original paper \cite{metalens}, we used metallic wires, half a wavelength long, as sub-wavelength resonators and we propose to explain the physics of this example. We focus on the resonant metalens introduced initially \cite{metalens} consisting of a square lattice of $N\times N$ ($N=20$) identical metallic wires, aligned along the vertical axis (defining the longitudinal $z$-direction). The length of the wires along $z$ is equal to $40 \textrm{ cm}$ and their diameter to $3\textrm{ mm}$. The period lattice in both transverse directions ($xy$ plane) is $1.2 \textrm{ cm}$, which is roughly equals to $\lambda/70$ for the first intrinsic resonance of a single wire.

The lens of interest here falls within the class of "wire medium". We define the latter as a uniaxial medium formed by a periodic lattice of conducting wires with small radii compared to the lattice periods and the wavelength. In microwave applications this medium is often seen as an interesting artificial dielectric, which presents the great advantage to have a negative effective permittivty \cite{PendryWireMedium}. Some attention to wire media has been paid in the realization of left-handed media as composite media made from lattices of long conducting wires and split ring resonators \cite{Smith}. Usually, quasistatic models for this medium are carried out for wave propagation perpendicular to the wires which show that, for electric-field polarization along the wires, the medium is characterized by a frequency-dependent effective dielectric constant. 

Belov and his colleagues \cite{belovTM,belov2,PREBelov} has studied this kind of medium in details, and especially the propgation along the wires. They showed that this medium can be described by means of an anisotropic dispersive dielectric tensor. Also, they demonstrated that the wire medium  supports  propagating  modes, so-called transmission line modes, which travel along the wires. During the whole article, we will focus only on these modes which explain our initial results \cite{metalens} and permit to give a simple explanation for our meaning. 

These modes which propagate in an adispersionless way in an infinite wire medium gives rise to dispersion when adding finiteness along the wires.  While in the original paper \cite{metalens} we insisted on the fact that the dispersion is a key issue for sub-wavelength focusing/imaging with our method, we will give a simplified model for the physical mechanism responsible for it and a discussion justifying our simplification is presented in appendix. We show why such a medium can be used to code a sub-wavelength wavefield into a temporal/frequencial signature thanks to its dispersion relation.

This will lead us to note the link between our work and Pendry's designers plasmons \cite{PendryPlasmons}, hence introducing the waves propagating in a finite length wire medium as new spoof plasmons. We will as well comment the very interesting effect which occurs at a specific frequency, known as canalization regime\cite{belov,belovCanalization}, at the light of our simplified approach. 

The other key issue for a resonant metalens is the far field radiation of the sub-wavelength features. Usually, a sub-wavelength wave field  is considered as evanescent, which forbids its propagation toward the far field. But, the notion of evanescent waves, which is responsible for the diffraction limit, is a mathematical formalism which only fits that of infinite interfaces. Using finite transverse dimensions and ending on the medium presented initially\cite{metalens}, we show that the sub-diffraction details of the latter actually contributes to the far-field radiation. 

Eventually, it is commonly admitted that their contribution to the total radiation is much weaker than those of propagating diffraction limited waves. Due to the resonant nature of the medium of interest, another physical mechanism which is well known from opticians intervene: the Purcell effect \cite{purcell}. This effect increases the coupling of the source with radiating modes,  and surprisingly it counterbalances the effect of the weak radiation of the sub-wavelength features.  

\section{The infinite wire medium: adispersionless propagation of TEM Bloch waves}
\begin{figure}
\includegraphics[width=8.5cm]{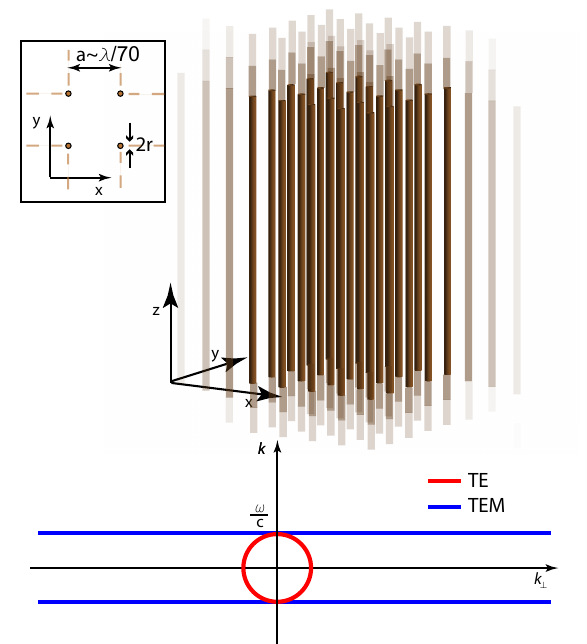}
\caption{\label{unbounded_wire_medium}  (top) The geometry of the infinite wire medium:  an infinite square array of parallel perfectly conducting thin wires. We focus on the frequency range where the lattice parameter is sub-wavelength. (bottom) Dispersion  curves  for  the TE and TEM types of Bloch waves supported by the medium. The TEM Bloch modes are always propagating with a flat dispersion curve. For the geometry under consideration ($r\ll a\ll\lambda$) the TM Bloch modes are always evanescent\cite{belov2}.}
\end{figure}

We consider an infinite "wire medium"  consisting of infinite Perfect Electric Conductors (PEC) wires aligned along the $z$-axis, forming in the $xy$-plane a two dimensional square lattice (Fig. \ref{unbounded_wire_medium}). We assume a thin wire approximation, ie. their radii are really small compared to the lattice constant. We are looking for electromagnetic fields $(\mathcal{E}; \mathcal{H})$ solutions of the following Maxwell problem and satisfying the boundary conditions:

\begin{eqnarray}
\vec{\nabla}\times\mathcal{E}&=&-\mu_0 \frac{\partial \mathcal{H}}{\partial t} \nonumber \\ 
\vec{\nabla}\times\mathcal{H}&=&\varepsilon_0 \frac{\partial \mathcal{E}}{\partial t}
\end{eqnarray}

\noindent where $\mu_0$, $\varepsilon_0$ are respectively the vacuum permeability and permittivity. Here we have assumed that the electromagnetic fields will propagate in vacuum only: the presence of metallic surfaces will be taken into account through boundary conditions only. Using the invariance of the medium along the $z$-axis and choosing a time dependence in $e^{-i \omega t}$, we define time-harmonic two-dimensional electric and magnetic fields $\mathbf{E}$ and $\mathbf{H}$ as:

\begin{eqnarray}
\mathcal{E}(\mathbf{x_\perp},z,t)&=&\Re e \left(\mathbf{E}(\mathbf{x_\perp})e^{i(k z-\omega t)}\right) \nonumber \\ 
\mathcal{H}(\mathbf{x_\perp},z,t)&=&\Re e \left(\mathbf{H}(\mathbf{x_\perp})e^{i(k z-\omega t)}\right)
\end{eqnarray}

\noindent where $\mathbf{x_\perp}$ stands for the coordinates in the $(xy)$ plane and $k$ is the propagation constant of waves along the $z$-axis. 
$\mathbf{E}$ and $\mathbf{H}$ are complex valued fields depending on two variables (coordinates $x$ and $y$) but still having three components along the three axes. The medium of interest is an uniaxial medium and we propose to decompose the electromagnetic fields with respect to the geometry of the medium: 

\begin{eqnarray}
\mathbf{E}(\mathbf{x_\perp})&=&\mathbf{E_\perp}(\mathbf{x_\perp}) + E_\parallel(\mathbf{x_\perp})\mathbf{e_z}\nonumber \\ 
\mathbf{H}(\mathbf{x_\perp})&=&\mathbf{H_\perp}(\mathbf{x_\perp}) + H_\parallel(\mathbf{x_\perp})\mathbf{e_z}
\end{eqnarray}

Eventually, due to the periodic nature of the medium in the transverse plane, the problem becomes the well known problem in solid state physics\cite{kittel} which consists in looking for Bloch waves solutions $U_{\mathbf{k_\perp}}$ (where $U$ stands for $\mathbf{E_\perp}$,  $\mathbf{H_\perp}$,  $E_\parallel$ or $H_\parallel$) that have the form:

\begin{equation}
U_{\mathbf{k_\perp}}(\mathbf{x_\perp})=e^{i\mathbf{k_\perp}.\mathbf{x_\perp}}\tilde{U}(\mathbf{x_\perp})
\end{equation}

\noindent where $\mathbf{k_\perp}$ is the so-called Bloch wave number (or quasi-momentum in solid state physics) and $\tilde{U}(\mathbf{x_\perp})$ is a function that satisfies the periodicity of the wire medium ($\tilde{U}(\mathbf{x_\perp})=\tilde{U}(\mathbf{x_\perp}+\mathbf{R})$ where $\mathbf{R}$ is a lattice vector). With these geometric considerations the Maxwell system reduces to:

\begin{eqnarray}\label{Maxwell}
i\omega\mu_0 \tilde{H}_\parallel\mathbf{e_z}&=&\left(i \mathbf{k}_\perp + \vec{\nabla}\right)\times\tilde{\mathbf{E}}_\perp \nonumber \\ 
i\omega\mu_0\tilde{\mathbf{H}}_\perp&=&\mathbf{e_z}\times\left(ik\tilde{\mathbf{E}}_\perp-\left(i \mathbf{k}_\perp + \vec{\nabla}\right)\tilde{E}_\parallel\right) \nonumber \\
-i\omega\varepsilon_0 \tilde{E}_\parallel\mathbf{e_z} &=&\left(i \mathbf{k}_\perp + \vec{\nabla}\right)\times\tilde{\mathbf{H}}_\perp\nonumber \\
-i\omega\varepsilon_0\tilde{\mathbf{E}}_\perp&=&\mathbf{e_z}\times\left(ik\tilde{\mathbf{H}}_\perp-\left(i \mathbf{k}_\perp + \vec{\nabla}\right)\tilde{H}_\parallel\right)
\end{eqnarray}

Then, we study independently the three types of Bloch modes that can exist with respect to the wires: the transverse magnetic (TM) where $\tilde{H}_\parallel=0$, the transverse electric (TE) where $\tilde{E}_\parallel=0$, and the transverse electromagnetic (TEM) modes where $\tilde{H}_\parallel=\tilde{E}_\parallel=0$. The final equation to be solved for each type of mode becomes:

\begin{eqnarray} 
\textrm{TM}:& \left(k_\perp{}^2+k{}^2-\left(\frac{\omega}{c}\right)^2\right)\tilde{E}_\parallel-2i(\mathbf{k}_\perp.\vec{\nabla})\tilde{E}_\parallel-\Delta\tilde{E}_\parallel=0 \nonumber\\
\textrm{TE}:& \left(k_\perp{}^2+k{}^2-\left(\frac{\omega}{c}\right)^2\right)\tilde{H}_\parallel-2i(\mathbf{k}_\perp.\vec{\nabla})\tilde{H}_\parallel-\Delta\tilde{H}_\parallel=0 \nonumber\\
\textrm{TEM}:& \hspace{.5cm}\left(k{}^2-(\frac{\omega}{c})^2\right)\tilde{\mathbf{E}}_\perp=\vec{0}
\end{eqnarray}

Solving the boundary value problem for TE and TEM modes is easy since they do not have any electric field component along the wires, and a dispersion law is easily obtainable from the previous equations. For the TM modes this boundary problem is a tedious work which has already been performed\cite{belovTM}. As a summary, the solutions of the Maxwell eigenproblem in the first Brillouin zone must satisfy the following dispersion relations (Fig. \ref{unbounded_wire_medium}):

\begin{eqnarray} 
\textrm{TM}&:& k_\perp{}^2+k{}^2 = \frac{\omega^2}{c^2}-k_p{}^2 \nonumber \\
\textrm{TE}&:& k_\perp{}^2+k{}^2 = \frac{\omega^2}{c^2} \nonumber \\
\textrm{TEM}&:& k{}^2 = \frac{\omega^2}{c^2}
\end{eqnarray}

\noindent where $k_p$ is the so-called plasma wave number, a parameter that depends on the lattice parameter $a$ and the wires radii $r$ and its expression can be found in the litterature\cite{belovTM,belov2}. 

From now, we focus on the range of angular frequencies $\omega$ where the associated vacuum wavelength is really higher ($> 10$) than the distance between two wires. In that case and in the thin wire approximation, the plasma wavenumber is greatly higher\cite{belovTM,belov2} than the free space one and the TM waves are always evanescent. The propagating TE modes along the $z$-direction are included inside the vacuum light cone. And the TEM Bloch modes are always propagating with a flat dispersion curve independent of the norm of the Bloch wave number $k_\perp$. 
Clearly, for the  large $k_\perp$  corresponding  to sub-wavelength features, the TE and TM modes are evanescent, while the TEM ones are propagating. We will only consider the propagating modes for this range, ie. the TEM Bloch modes. Thus, the fields inside the structure for a given $k_\perp$ can be described by a unique scalar potential $\tilde\phi(x_\perp)$:

\begin{eqnarray} \label{potential}
\mathbf{E}_{\mathbf{k_\perp}}(\mathbf{x_\perp})&=&-\vec{\nabla}e^{i\mathbf{k_\perp}.\mathbf{x_\perp}}\tilde{\phi}(\mathbf{x_\perp})\nonumber \\ 
\mathbf{H}_{\mathbf{k_\perp}}(\mathbf{x_\perp})&=&\frac{1}{\mu_0c}\mathbf{e_z}\times\vec{\nabla}e^{i\mathbf{k_\perp}.\mathbf{x_\perp}}\tilde{\phi}(\mathbf{x_\perp})
\end{eqnarray}

\noindent $\tilde\phi(\mathbf{x_\perp})$ must satisfy the periodicity of the medium as well as the boundary conditions at metallic surfaces. This potential can be seen as a spatial envelope that permits to verify the constraints, and it multiplies a plane wave, as a consequence of the Bloch theorem. 

During the rest of this article, we will always neglect the high diffraction orders due to this potential and consider it as a constant potential. We will always talk about plane waves instead of Bloch waves neglecting the Bloch potential. One important consequence of this hypothesis is that  we can arbitrarily change the coordinate axes without lack of generality, and especially we will often choose $\mathbf{k}_\perp=k_\perp\mathbf{e_x}$. 

Another interesting result concerning the TEM Bloch modes is that the group velocity $\vec{v}_g=\vec{\nabla}_\mathbf{k}(\omega)=c\,\mathbf{e_z}$ is constant and oriented along the $z$-axis. Therefore any spatial wavepacket in a transverse plane propagates without distortion along the $z$-axis and one can use this property for guiding and imaging subwavelength features \cite{shvets}.

\section{The finite wire medium along the wires axis: a dispersive medium}

\subsection{Snell's law for a wire medium-air interface}

Unfortunately, the infinitely extended wire medium cannot exist and introducing a boundary condition leads to interesting results. Here, we consider a TEM Bloch wave with transverse wave number $\mathbf{k_\perp}=k_\perp\mathbf{e_x}$ traveling in the half space $z<L/2$ filled out with a wire medium. This Bloch wave encounters a boundary with vacuum at $z=L/2$. The incident angle $\theta_i$ is defined by $k_\perp/k$. This wave generates a reflected wave with angle $\theta_r$ and a transmitted wave with angle $\theta_t$. The first step consists in finding an equivalent of the Snell's law for this interface. Considering the conservation of the tangential component of the wave number at the interface (here $k_\perp$) and introducing the dispersion relation on both sides of the equality we obtain:

\begin{eqnarray} \label{snell}
\theta_r&=&-\theta_i \nonumber\\
\tan\theta_i&=&\sin\theta_t
\end{eqnarray} 

\noindent From this refraction's law the critical incident angle $\theta_c=\pi/4$ appears: when the incident angle is higher than $\theta_c$ (ie. for evanescent waves in the air region) a total internal reflection is observable. 

In order to simplify the calculation for continuity of tangential components, as explained in the previous part, we will neglect the evanescent TM wave that is generated in the wire medium at the reflection and the high diffraction orders. A discussion concerning those approximations is presented in appendix. Then, according to equation (\ref{potential}), the electric fields on both sides of the interface write:

\begin{eqnarray}
\mathbf{E}^{(i)}&=&E_0 e^{i(k_\perp x+k(z-L/2))}\left( \begin{array}{c} 1\\0\\0 \end{array} \right) \nonumber \\
\mathbf{E}^{(r)}&=&r E_0  e^{i(k_\perp x-k(z-L/2))}\left( \begin{array}{c} 1\\0\\0 \end{array} \right) \\
\mathbf{E}^{(t)}&=&t E_0 e^{i(k_\perp x+\kappa(z-L/2))}\left( \begin{array}{c} \cos\theta_t\\0\\ -\sin\theta_t \end{array} \right)   \nonumber
\end{eqnarray}

And the associated magnetic fields write:

\begin{eqnarray}
\mathbf{H}^{(i)}&=&\frac{E_0}{\mu_0c} e^{i(k_\perp x+k(z-L/2))}\left( \begin{array}{c} 0\\1\\0 \end{array} \right) \nonumber \\
\mathbf{H}^{(r)}&=&\frac{r E_0}{\mu_0c}  e^{i(k_\perp x-k(z-L/2))}\left( \begin{array}{c} 0\\-1\\0 \end{array} \right) \nonumber \\
\mathbf{H}^{(t)}&=&\frac{t E_0}{\mu_0c} e^{i(k_\perp x+\kappa(z-L/2))}\left( \begin{array}{c} 0\\1\\ 0\end{array} \right)  
\end{eqnarray}

\noindent where $k=\frac{\omega}{c}$ stands for the free space wave number, and $\kappa=\sqrt{k^2-k_\perp{}^2}$ if $\theta_i\leq\theta_c$ or $\kappa=i\sqrt{k_\perp{}^2-k^2}$ if the total internal reflection occurs.
Applying the continuity of the tangential components of the fields at the interface and introducing the refraction law of equation (\ref{snell}) yields to the following reflection coefficient:

\begin{equation} \label{r}
r=-\frac{k-\kappa}{k+\kappa} 
\end{equation}

Interestingly, in the peculiar case of total internal reflection, $r$ becomes a complex coefficient with unity norm, and the reflection induces a phase shift:

\begin{equation}\label{phi}
\varphi(k_\perp,k)=\pi-2\arctan\left(\frac{\sqrt{k_\perp{}^2-k^2}}{k}\right)
\end{equation}

\noindent This phase shift is a well-known phenomenon for total internal reflection and one of its consequences is the Goos-H\"anchen shift \cite{Goos} along the $x$-direction. Here, this phase shift has an importance when adding a second boundary at $z=-L/2$. Now, the  wire medium with finite dimension consists of wires aligned along the $z$-direction with the same length $L$ (Fig. \ref{dispersion}). With this new constraint the field inside the medium experiences a Fabry-Perot like resonance, or equivalently a guided propagation along $x$-axis, when the optical path length $\delta$ for a round trip satisfies:

\begin{equation}\label{delta}
\delta=2kL+2\varphi(k_\perp,k)=2n\pi
\end{equation} 

\noindent where $n$ is an integer. With equation (\ref{phi}) and (\ref{delta}) we can write a dispersion relation  for the guided waves inside the bounded wire medium for the first resonance ($n$=1):

\begin{equation}\label{dispersion1}
\tan\left(k\frac{L}{2}\right)=\frac{\displaystyle\sqrt{k_\perp{}^2-k^2}}{k}
\end{equation}

This implicit theoretical dispersion relation is plotted in terms of the first Fabry-Perot frequency $f_m$  versus $k_\perp$ (both normalized to  $f_0$ and $k_0$ corresponding to an ideal Fabry-Perot of length $\lambda/2$) in Figure \ref{dispersion}. The asymptotes of the light line at low frequencies and the intrinsic resonance frequency at large values of $k_\perp$ are shown. 

Rigorously, the TM evanescent waves as well as the high diffraction orders (and thus the TE diffracted waves inside the wire medium) induced at the reflection adds small corrections to the phase shift. This would lead to a more complicated dispersion law that depends on the wire medium parameters. For simplicity and because this dispersion is in a sufficiently good agreement with simulation results (Fig. \ref{dispersion_with_TM}), we have preferred not to introduce those waves. More sophisticated dispersion laws are presented in the appendix. We believe our calculation that only takes into account the evanescent field at reflection is the easiest way to understand why the $z$-bounded wire medium introduces the dispersion. Because this dispersion is a key issue for the concept of resonant metalens, we have chosen to make those approximations for the scope of this article, but obviously one can add small corrections to this law.

\subsection{Fields profiles inside the medium}

\begin{figure}
\includegraphics[width=8.5cm]{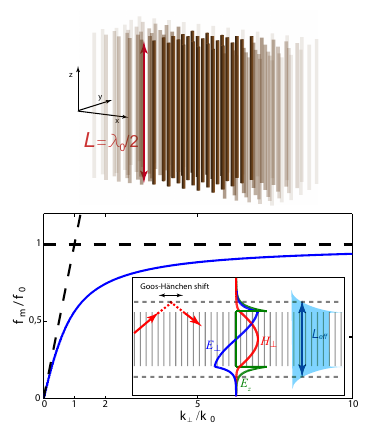}
\caption{\label{dispersion} (top) The geometry of the infinite wire medium with finite length along the wires. (bottom) The dispersion relation of the first resonance frequencies $f_m$ (normalized to the original Fabry-Perot frequency $f_0$) versus the transverse Bloch wave number $k_\perp$ (normalized to the original Fabry-Perot wave number $k_0$) from equation (\ref{dispersion1}),(\ref{dispersion2}) or (\ref{dispersion3}).  The asymptotes of the light cone  and the original resonance frequency are shown (dashed-lines). The inset shows how the three ways of calculation (reflection coefficient, guided waves expression and effective Fabry-Perot length) leads to the same result.}
\end{figure}

In order to describe precisely these propagating waves inside the wire medium we consider that the region between $z=\pm L/2$ is filled out with a wire medium (Fig. \ref{dispersion}). We know that the field inside this region is TEM with respect to the wire but the invariance along $z$ is broken. Still considering the first order of diffraction, neglecting TM evanescent waves, and choosing the $x$-axis in order to have $\mathbf{k_\perp}=k_\perp\mathbf{e_x}$, we can express the fields inside the wire medium as:

\begin{eqnarray}
\mathbf{E}_\textrm{TEM}&=&e^{i\mathbf{k_\perp}.\mathbf{x_\perp}}\left(\begin{array}{c} E_0(z)\\0\\0 \end{array} \right) \nonumber \\ 
\mathbf{H}_\textrm{TEM}&=&\frac{1}{i\mu_0\omega}e^{i\mathbf{k_\perp}.\mathbf{x_\perp}}\left(\begin{array}{c} 0\\ \textrm{d} E_0/\textrm{d} z\\0 \end{array} \right)
\end{eqnarray}

And, from the propagation equation (\ref{Maxwell}) of the TEM Bloch modes inside the wire medium it follows:

\begin{equation}
E_{0}(z)=C_{1}\sin\left(\frac{\omega z}{c}\right)+C_{2}\cos\left(\frac{\omega z}{c}\right)
\end{equation}

In the air region on both sides of the wire medium the electromagnetic fields decompose onto a plane wave that we can write using the previous notations:

\begin{eqnarray}
\mathbf{E}_\textrm{vac}&=&E_{1,2} e^{i(\mathbf{k_\perp}.\mathbf{x_\perp} \pm \kappa (z\mp\frac{L}{2}))} \left(\begin{array}{c} \pm \kappa/k \\0\\  -k_\perp/k \end{array} \right)\nonumber \\ 
\mathbf{H}_\textrm{vac}&=&\frac{E_{1,2}}{i\mu_0 c}e^{i(\mathbf{k_\perp}.\mathbf{x_\perp} \pm \kappa (z\mp \frac{L}{2}))} \left(\begin{array}{c} 0 \\1\\ 0 \end{array} \right)
\end{eqnarray}

Then applying the continuity of the tangential fields at the interface we reach the matrix form equation:

\begin{equation}
\begin{pmatrix}
\sin\left(kL/2\right) & \cos\left(kL/2\right) & -\kappa/k & 0 \\
-\cos\left(kL/2\right) & \sin\left(kL/2\right) & -i & 0  \\
-\sin\left(kL/2\right) & \cos\left(kL/2\right) & 0 & \kappa/k \\
-\cos\left(kL/2\right) & -\sin\left(kL/2\right) & 0 & -i  \\
\end{pmatrix}
\begin{pmatrix}
C_1 \\ C_2 \\ E_1 \\ E_2
\end{pmatrix}
=\vec{0}
\end{equation}

The system has a non-zero solution only if its matrix determinant is null. The first solution of this cancellation gives another expression of the dispersion relation of equation (\ref{dispersion1}):

\begin{equation} \label{dispersion2}
\cos\left(k\frac{L}{2}\right)=\frac{k}{k_\perp}
\end{equation} 

This expression is actually the same as the one presented previously in equation (\ref{dispersion1}), and one can develop the sine function from its cosine and recover the tangent function. From this expression, we can eventually express all of the electromagnetic components. In the wire medium region it gives:

\begin{eqnarray}
\mathbf{E}_\textrm{TEM}&=&E_0 \sin\left(k z\right) e^{i(\mathbf{k_\perp}.\mathbf{x_\perp})} \mathbf{e_x} \nonumber \\
\mathbf{H}_\textrm{TEM}&=&-\frac{i E_0}{\mu_0 c}\cos\left(k z\right) e^{i(\mathbf{k_\perp}.\mathbf{x_\perp})} \mathbf{e_y}
\end{eqnarray}

One can notice that for the first resonance the electric field is maximum at both ends of the wire medium, while the magnetic field is maximum at half the wires length, and is near zero at both ends (fig. \ref{dispersion}). And in the vacuum regions for high Bloch wave numbers (ie. the waves are evanescent in air) it gives:

\begin{eqnarray}
\mathbf{E}_\textrm{vac}&=& E_0 e^{i\mathbf{k_\perp}.\mathbf{x_\perp}}e^{\mp \sqrt{k_\perp{}^2-k^2} (z\mp \frac{L}{2})} \left( \begin{array}{c} \pm\sqrt{1-k^2/k_\perp{}^2}\\0\\i \end{array} \right)
  \nonumber \\
\mathbf{H}_\textrm{vac}&=&  \frac{i E_0}{\mu_0 c}\frac{k}{k_\perp}  e^{i\mathbf{k_\perp}.\mathbf{x_\perp}}e^{\mp \sqrt{k_\perp{}^2-k^2} (z\mp \frac{L}{2})}\mathbf{e_y}
\end{eqnarray}

\subsection{Energy density considerations}

However, we can use a more intuitive approach to determine the dispersion law. In fact, depending on the transverse wavevector $k_\perp$, the Bloch waves penetrate in air up to a distance whose inverse is the quantity previously introduced $\sqrt{{k_\perp}^2-k{}^2}$. Again, a more sophisticated model would add correction to this penetration depth taking into account high diffraction orders and TM waves (see appendix). Each propagating mode in the structure experiences the Fabry-Perot resonance, and the length of the cavity is not simply the length $L$ of the wire medium, because of the penetration of the modes inside air. A common way to describe the effective length of the real cavity is to calculate the quantity:
	
\begin{equation}
L_{\textit{eff}}=\frac{\displaystyle\left[\int_{-\infty}^{\infty}u(x,y,z)\textrm{d}z\right]^2}{\displaystyle\int_{-\infty}^{\infty}\left[u(x,y,z)\right]^2\textrm{d}z}
\end{equation}
					 
where $u(x,y,z)$ is the energy density of a given mode. Using the fact that the modes are TEM inside the structure, the density of energy inside the structure is constant along the $z$-direction, from $-L/2$ to $L/2$. For the sake of simplicity, we set this value to $1$, which does not hamper our description since this value simplifies in $L_{\textit{eff}}$. The energy density decreases exponentially at the interfaces $z=-L/2$ and $z=L/2$ and taking into account the three non-zero components of the fields, the calculus simply gives:

\begin{equation}
L_{\textit{eff}}=L+\frac{2}{\sqrt{{k_\perp}^2-k{}^2}}
\end{equation}
 
This expression has merit to show explicitly the increase of the Fabry-Perot cavity length due to penetration depths at interfaces. We finally consider that the resonance condition is obtained if $L_{\textit{eff}}=n \lambda/2$ for the $n$th Fabry-Perot like condition, which leads after rearrangement to the dispersion relation initially presented \cite{metalens} for  the first Fabry-Perot resonance:

\begin{equation}\label{dispersion3}
\frac{f_0}{f}=1+\frac{2}{\pi}\left[\left(\frac{k_\perp}{k_0}\right)^2-\left(\frac{f}{f_0}\right)^2\right]^{-\frac{1}{2}}
\end{equation}

\noindent where $k_0$ and $f_0$ stands for the original first Fabry-Perot like resonance, namely $k_0=\pi/L$ and $f_0=c/2L$. 

\subsection{Link with designer plasmons}
 
The dispersion curve deserves some comments. Indeed, this relation is similar to the ones characterizing surface plasmons polaritons (SPPs) and we have demonstrated the existence of guided modes inside the wire medium (with finite length) with high wavenumbers. But, at low frequencies, where the limit of perfect conductor is applicable, it is well known that perfect metals do not support electromagnetic surface modes, forbidding the existence of SPPs. 

However, it has been demonstrated theoretically \cite{PendryPlasmons} and experimentally \cite{Hibbins} that bound electromagnetic surface waves mimicking SPPs can be sustained even by a perfect conductor, provided that its surface is periodically corrugated. Two distinct geometries of modulation of the flat perfectly conducting surface have been introduced \cite{Vidal}: a one-dimensional array of grooves and a two-dimensional hole array. The frequencies of the supported modes scale with the geometrical size of the corrugations in the perfect conductor approximation. Maier and his colleagues \cite{Maier} also demonstrated the existence of pseudo surface waves in the low frequency regime: structures  consisting  of  closely spaced metal rods in one dimension can guide electromagnetic waves. In the present case, the same configuration has been investigated but in the two dimensional case, and we have demonstrated the same guiding properties. 

Interestingly, the equation (\ref{dispersion1}) is actually the same dispersion law as the one obtained for the spoof plasmons \cite{Vidal}. Again, in both cases the high diffraction orders and the mode conversions are neglected when calculating the dispersion law, and this is the reason why the two laws are the same. Because of the matching of the dispersion laws at the first order, one can imagine to couple the two kinds of structure. Similarly, the thickness of the layer that supports these plasmon-like waves scales with the operating frequencies since the height of the wires is near $\lambda/2$. But, here, only the length of the wires has an influence on the resonance frequency. It is not the case for the array of grooves where the groove's width and the distance between two of them have an impact on the dispersion curve. In the array of pinholes, the fundamental mode inside a hole is considered which implies a dependence of the hole dimensions in the dispersion. Here, the distance between wires as well as wires diameters do not influence the dispersion law and the guiding properties of the medium do not seem to be significantly altered when wires are displaced from their original positions. Another interesting property is that the wire medium is essentially made out of vacuum, thus  losses induced by metallic surfaces are less important than in the corrugated  surfaces or in the surface with pinholes.

\subsection{Link with canalization regime}
Eventually, we would like to add comments concerning the results obtained by Belov and his colleagues \cite{belov}. They have used the wire medium for performing a pixel to pixel imaging with a resolution thinner than the diffraction limit. The same near field  lens has also recently reached the near infra-red spectrum \cite{Casse} and sub-wavelength imaging has been proved experimentally using gold nano-wires. In the two cases, the approach is monochromatic and the operating frequency of their lens corresponds to the intrinsic resonance of a single wire, namely $f_0$. At this exact frequency the field distribution at an interface is recovered at the other interface, and by scanning near field of one interface they perform sub-wavelength imaging. We would like to prove that their results corresponding to the canalization regime are consistent with the ones developed here. From previous calculations, introducing equation (\ref{dispersion1}) into equation (\ref{r}) leads to the following reflection coefficient:

\begin{equation}
r(f)=-\frac{1-i\tan(\pi f L/c)}{1+i\tan(\pi f L/c)}
\end{equation}

Therefore, at the intrinsic resonant frequency of a single wire $f_0$ this reflection coefficient is equal to $-1$ and does not depend on the Bloch wavenumber $k_\perp$ since the dispersion relation has no solution at this frequency. Because this coefficient is real, the reflection does not imply any phase shift and especially the Goos-H\"anchen shift is null. As we have seen before, the group velocity for the TEM Bloch waves inside the wire medium is oriented along the wire axis, so if there is no Goos-H\"anchen shift at the reflection, there is no propagation in the transverse plane. At $f_0$, whatever the Bloch wavenumber $k_\perp$ considered the reflected wave propagates along the wires as well as the incident one, hence giving rise to the so called canalization regime. This regime implies that the electric field has the same profile at both interfaces. When placing a sub-wavelength source at an interface they managed to image it half a wavelength away at the other interface with a resolution far better than the diffraction limit\cite{belov,PREBelov}.

\subsection{Coding of sub-wavelength details}

The dispersion is a key issue for the concept of resonant metalens we introduced \cite{metalens}. The importance of this dispersion manifests when a sub-wavelength source is placed near one of the two interfaces. Actually, the two interfaces correspond to the object plane of the lens and the lens permits to obtain the field profile inside one of these transverse planes. This profile decomposes with a unique set of amplitude and phase on the modes supported by the wire medium. Typically, it can be expressed in terms on $k_\perp$ since it corresponds to a Fourier transform:

\begin{equation} \label{resolution}
E(\mathbf{x_\perp}) = \iint \tilde{E}(\mathbf{k_\perp}) e^{i\mathbf{k_\perp.x_\perp}} d\mathbf{k_\perp} 
\end{equation}

\noindent where in the present case the integration domain is limited to the first Brillouin zone of the lattice, that is to say when one of the component of the Bloch wavenumber reaches $\pi/a$.

 This formalism makes sense when considering broadband illuminations. Interestingly, from previous calculations and ignoring the degeneracy, a given Bloch wavenumber $k_\perp$ is associated to its own angular frequency $\omega$. For the vacuum dispersion relations this is also the case but here $k_\perp$ can be associated to a spatial scale that is deeply sub-wavelength compared to the associated vacuum wavelength. This property permits to transform the function $\tilde{E}(\mathbf{k_\perp})$ into a function that depends only on frequency: this is the frequency coding of the sub-wavelength details. This property means that, when the medium is illuminated from one of the two object planes, registering the frequency spectrum of a wave packet gives the spatial Fourier transform of the source. The spatial features of the source can now be expressed in terms of $\omega$:

\begin{equation}  
\tilde{E}(\mathbf{k_\perp}) = \mathcal{F}(\omega) 
\end{equation}

Eventually, it is well known that time and frequency are two distinct representations of the same phenomena. A broadband illumination means the emission of a short pulse, and the frequency signature $\mathcal{F}(\omega)$ for the sub-wavelength features means a temporal one $f(t)$. The imaging procedure associated to this concept is the scope of the next article, but at this time one can easily understand that the resolution must be limited by the distance between two wires. By exploiting the dispersion of resonant modes with broadband illuminations we can record the sub-wavelength details of an object. In the next part we will see that this information can be carried toward far field.

\section{A wire medium with finite transverse dimensions: far field radiation of sub-wavelength features}

\begin{figure}
\begin{center}
\includegraphics[width=8.5cm]{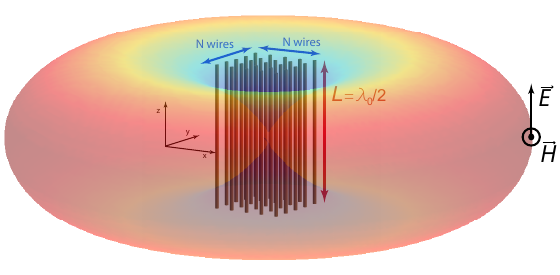}
\caption{\label{bounded_wire_medium} The geometry of the wire medium with finite dimensions in the 3 spatial directions. Due to the finiteness this medium radiates field toward the far field and superimposed on the medium we have represented a sketch of a radiation pattern for the first resonance. This is just an example since the next figure shows that the directivity pattern in the $(xy)$ plane can be different.}
\end{center}
\end{figure}

Until now, everything happened inside or in the near field of the wire medium. Now, we consider the previous wire medium but with finite dimensions in both transverse directions (Fig. \ref{bounded_wire_medium}). Introducing this finiteness has a first consequence which is the quantization of the transverse wavenumbers. In solid state physics, the quantization is done by introducing Born–Von Karman boundary conditions \cite{kittel} which consists in cycling a crystal. When studying crystals the number of atoms considered is large enough to allow this approximation. In the present case, the number of wires in both transverse directions is only $20$ and this boundary condition is not well suited. 

But, introducing finiteness results in the same quantization effect. At each interface, the plasmon-like wave traveling inside the wire medium encounters a boundary with vacuum and gives birth to a reflected one. The analytical expression of the reflection coefficient is not easy to calculate since the field in the air region is the superposition of many plane waves due to the finiteness along the wires.  Then, the interferences inside the wire medium are constructive only for a discrete spectrum of Bloch wavenumbers which characterize the stationary eigenmodes of the system. Thus, due to the finiteness of the array of wires, the transverse Bloch wave numbers are quantified: $\mathbf{k_\perp}=\frac{\pi}{D}(m.\mathbf{e_x}+n.\mathbf{e_y})$, with integers $(m,n)\in[\![0;N-1]\!]^2$ and $D$ the size of the medium in the $x$ and $y$ directions, $D=a(N-1)$. Again, from the previous section, it is important to keep in mind that each $k_{\perp(m,n)}$ is associated to its own resonant frequency $f_{(m,n)}$, ensuring a frequency (or equivalently a time) coding of spatial features.

The second consequence of these interfaces is a conversion from the plasmon-like modes to propagating waves in the free space region. Intuitively, with an approach of coupled resonators, one easily understands that each wire generates a monopolar radiation (for the first resonance) which is essentially $z$-polarized. The wire medium can therefore give rise to a superposition of monopolar radiations. With the approach we used until now, the simplest way to explain the radiation comes from the fields profiles inside the wire medium (inset of figure \ref{dispersion}). We have already seen that the fields cannot exit the structure from the top or the bottom interfaces because the waves are evanescent. The only way to escape the wire medium is from lateral interfaces. Since the $x$ and $y$ components of the electric field present an odd profile for the first Fabry-Perot resonance, they cannot contribute to far field radiation.
While the $z$-component of the electric field can be modeled by two small dipoles $\vec{p}_{1,2}$ positioned at $z=\pm L/2$. The field radiated by the superposition of these two infinitesimal dipoles writes, in the dipolar approximation and in spherical coordinates:

\begin{equation}
\mathbf{E}=\frac{\mu_0\omega^2p_0\sin\Phi}{4\pi r} e^{i k r}\left(e^{i\frac{kL}{2}\sin\Phi}+e^{i(-\frac{kL}{2}\sin\Phi+\varphi)}\right)\mathbf{e_\Phi}
\end{equation}

\noindent where $\Phi$ corresponds to the zenith angle, and $\varphi$ stands for the phase difference between the two dipoles. Depending on the parity of the Fabry-Perot resonance order the two dipoles are in phase or out of phase ($\varphi=0\textrm{ or }\pi$). Typically, for the first Fabry-Perot resonance, or for the field profile presented in figure \ref{dispersion}, the two dipoles are in phase, ie. $\varphi=0$. Thus, for the first Fabry-Perot resonance, the radiated amplitude is maximum in the plane $z=0$ and it can be studied as a two dimensional problem.

Now we consider that we have such small dipoles on both interfaces of the wire medium with a 2D spatial distribution given by the $z$-component of the TEM Bloch mode at the interfaces. We perform the projection of this mode onto the basis of 2D free space plane waves. The $E_z$ field at the interfaces of the structure has the periodicity given by the Bloch wavenumber and has the form:

\begin{equation}
E_z^{\mathbf{k_\perp}}\propto \sin\!\left(k_{\perp_x}(x-\frac{D}{2})\right)\sin\!\left(k_{\perp_y}(y-\frac{D}{2})\right)\Pi_D(x,y)
\end{equation}
 
where  $\Pi_D$ is the 2D rectangle function of width $D$.

\begin{figure}
\begin{center}
\includegraphics[width=8.5cm]{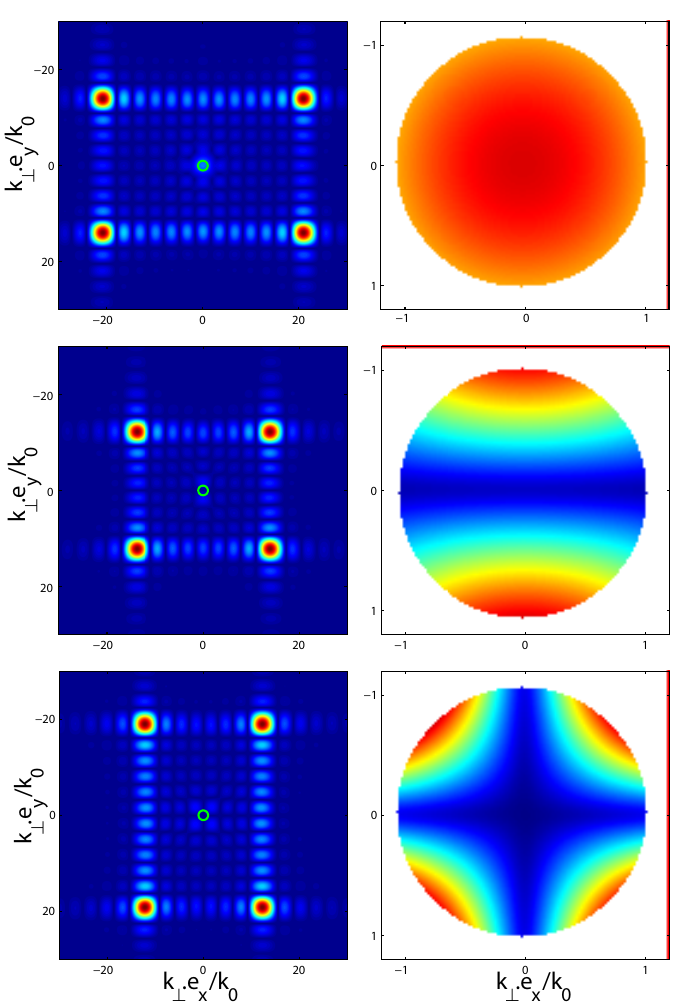}
\caption{\label{sinc} The two dimensional projection of three distinct Bloch eigenmodes onto the free space waves. The left colormaps show that the finiteness of the wire medium in the transverse plane lead to a superposition of 4 sinc functions. The right ones is the projection of the left ones onto the light cone: when considering object with finite dimensions a certain amount of energy can reach the far field. These colormaps also show that this representation gives information about the directivity pattern of an eigenmode. Depending on its $k_\perp$ monopolar, dipolar or quadrupolar radiation can occur.}
\end{center}
\end{figure}

	In the Fourier domain, this expression transforms into a sum of four sinc functions positioned at the corners of a rectangle of width ${2 k_{\perp_x}}$ and length ${2 k_{\perp_y}}$ (Fig. \ref{sinc}). Such a representation contains all information about the propagating fields generated by the sub-wavelength modes. The intersection of the latter with a disk of radius $\omega/c$ (green line in figure \ref{sinc}) gives the efficiency of the radiation as well as the directivity pattern. 	
	
	 Averaging this intersection along the disk resides in an estimation of the radiated energy in the far field for a given mode. The full calculation of the radiated energy resides in a projection onto a set of Hankel functions and is quite hard to perform, but an estimation of this quantity can be done. Due to the fact that we sum 4 sinc functions whose operands are proportional to $D k_{\perp_{x,y}}$, and due to the fact that the transverse dimension is lower than the operating wavelength, the radiation's efficiency in energy is roughly equal to $1/(16\pi).(DL k_{\perp_x} k_{\perp_y})^2$: the smaller the transverse wavenumber the higher the radiated amplitude. Fortunately, we will see later that the Purcell effect compensates for this low radiation efficiency. 
	
	In the same time, the spectral representation of the modes gives information about the directivity pattern of the radiated field. The intersection between this spectrum and the circle of radius $k$  directly gives the radiation pattern of the modes. A point on the circle with coordinates ($\cos\theta$,$\sin\theta$) permits to obtain the radiated amplitude of the mode in the direction defined by the angle $\theta$. The directivity pattern is therefore fully described by this representation. It is important to keep in mind that this representation is in the $(xy)$-plane and that the radiation in a plane containing $(Oz)$ is like the one of a single dipole antenna. To underline this aspect we represent the projection of the previous $k$-space spectra onto this circle (Fig. \ref{sinc}).
	
Depending on the mode parity, or equivalently of the parity of $m$ and $n$, the Fourier spectrum has to keep the same parity. For a given direction in the transverse plane, if the integer describing the wave number component is odd, the spectrum is even: it results in a null a value for the spectrum at the origin. On the contrary, an even integer does not impose a null value at the origin, and it does not cancel over because in the present case the transverse dimension $D$ is lower than half a wavelength. Taking into account the two directions it results in four directivity pattern possibilities: 
\begin{itemize}
\item the odd-odd modes give monopolar radiation;
\item the odd-even (resp. even-odd) modes give a dipolar along $x$ (resp. $y$) radiation;
\item the even-even modes give a quadrupolar radiation.
\end{itemize}

Therefore the resonant metalens possesses four spatial degrees of freedom, representing as many information channels exploitable for imaging and focusing \cite{Lemoult}. At a frequency where some modes are degenerate, registering far field in different directions allows a discrimination between modes, by the way increasing the number of subwavelength information registered.

\begin{figure}
\begin{center}
\includegraphics[width=8.5cm]{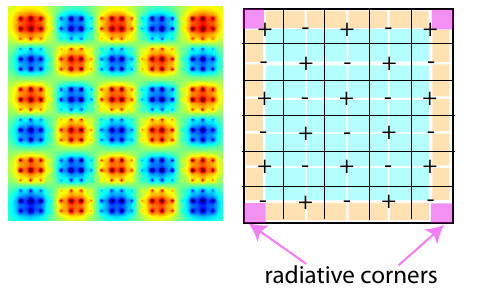}
\caption{\label{corners} Illustration of corner modes radiation. (left) The $z$-component of the electric field profile for a given Bloch eigenmode at one of the two interfaces $z=\pm L/2$. (right) Decomposition onto secondary sources demonstrating that the radiation is due to the corners. In this case one easily understand that a dipolar radiation is obtained.}
\end{center}
\end{figure}

This type of directivity patterns can also be interpreted in terms of corner radiation\cite{Lesueur,Fahy} which have been introduced in acoustics. In calculating radiation from vibrating square plates, a specific kind of modes, sometimes termed corner modes for the reason that only the corner quarter-cells contribute significantly to the radiation, have been described. Those modes satisfy the condition $k<m\pi/D$ and $k<n\pi/D$. These modes correspond to subsonic modes inside plates, analogue to the Bloch eigenmodes considered here. If we decompose the field onto subwavelength secondary sources, most of those sources tend to cancel one another. By grouping secondary sources onto dipoles or quadrupoles  (figure \ref{corners}), it is shown that the effective radiating area is limited to the four corners of the square, and the type of directivity pattern is easily obtained. 

\section{Purcell effect}
The idea we introduced in the original paper \cite{metalens} was to use the radiation of these subwavelength features to image an object placed near the resonant medium. Thus the last issue to be resolved is the coupling between the object and the lens itself. Spontaneous emission is not only an intrinsic property of the atoms or molecules in a specific material but is also governed by the electromagnetic properties of the host. In the 1950s, E.M. Purcell \cite{purcell} demonstrated that the rate of spontaneous emission for a light source depends on its environment: it can be increased when placing the source in a resonant cavity. This phenomenon is nowadays well known as "Purcell effect".

 Later, E. Yablonovitch \cite{Yablonovitch} predicted theoretically that a "photonic" material  can control the rate of radiative recombination of an embedded light source. This led to the huge domain of photonic bandgap materials where this effect is used for inhibiting the spontaneous emission of light. 
 
 Also, many theoretical works concern the effect of an absorbing dielectric on spontaneous emission  and  level  shifts  of  an  embedded  atom, using a Green's function approach \cite{Lagendijk,Fleischhauer}, or using a macroscopic Lorentz's cavity model \cite{Barnett}, or using equations of  motion  for  a  collection  of  two-level  atoms  embedded  in  a  dielectric  host medium \cite{Berman}. 
 As an atom in a dielectric is surrounded by many other two level resonators, these works are similar to the concept of resonant metalens we study here. Interestingly, we have already discussed the frequency shift in the case of the resonant metalens and here we discuss the increase of the spontaneous emission rate. Namely, in the resonant metalens case, the Purcell effect implies an increase of the coupling between the source and the wire medium since it acts as a cavity surrounding the source and modifies the so-called local density of states (LDOS).

The computation of the eigenmodes readily yields the quality factor of the modes resonance, a factor influencing the spontaneous emission. From the previous calculation, and in the deeply sub-wavelength limit, we showed that the radiation's efficiency is roughly proportional to  $(DL k_{\perp_x}k_{\perp_y})^2$: the lower the transverse wave number, the higher its radiated energy. In other words the resonance quality factor,  which is the ratio between the energy stored in the resonant medium and the radiated energy, grows as $(DL k_{\perp_x}k_{\perp_y})^2$. As stated by {E.M.} Purcell \cite{purcell},  it results in a better rate of spontaneous emissions for the highest transverse wave numbers. 

   This effect manifests itself by a better coupling between the source in front of the lens and especially for the most sub-wavelength modes. The resonant nature of the modes matches the impedance of the small electric dipole, or equivalently, the Purcell effect compensates for the weak radiation of the deep sub-wavelength modes. As a summary, we have seen that the Purcell effect implies a rate of spontaneous emission proportional to the quantity $(D k_\perp)^2$, while the radiation efficiency of a given mode is proportional to the inverse of this quantity. Overall, the two effects counterbalance themselves ensuring that all of the subwavelength details radiate the same amount of energy. And the higher the Bloch wavenumbers, the thinner the bandwidth carrying the energy.

\begin{figure}
\includegraphics[width=8.5cm]{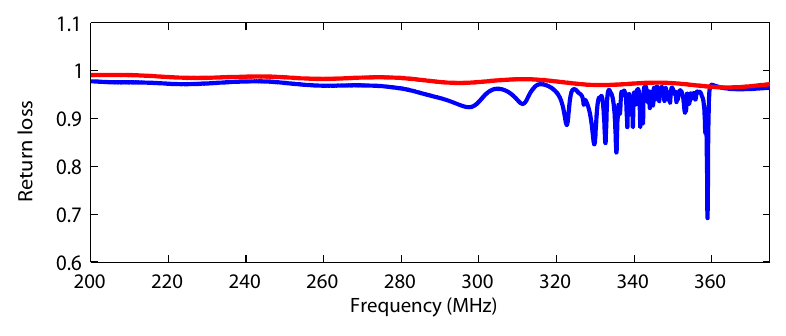}
\caption{\label{S11} Measured return loss for a small dipole with (blue) and without (red) the wire medium. Due to the resonant nature of the wire medium a completely reactive dipole becomes radiative, and this efficiency grows with the resonance Q factor, and thus with frequency here. The discrete nature of the blue curve still demonstrates that the subwavelength features are coded in the field spectrum.}
\end{figure}

On a temporal point of view, the $Q$ factor is the opposite of the attenuation time of the system: a higher quality factor implies a lower damping, and so high $Q$ modes oscillate for longer times.  This is why in the original paper of metalenses \cite{metalens} we presented a curve of the lifetimes of the modes inside the structure versus their Bloch wavenumbers $k_\perp$. Here, we have explained why the data were growing with $k_\perp$. The linear fit initially obtained results from the short time accessible for the time domain simulation compared to the extremely high $Q$ factor attainable.

In order to seal the validity of this physical description, we measured the return loss of a small dipole with a network analyzer (Agilent N5230C) with and without the presence of the lens (Fig. \ref{S11}). The return loss is the ratio of the power of the outgoing signal to the power of the signal reflected back from the small dipole. If the return loss is equal to 1, the whole input power is reflected back meaning that there is no radiated field. While when this quantity decreases, some amount of energy has been transmitted by the dipole: this energy corresponds to the radiated energy.  Without the lens, it results in a flat curve in amplitude demonstrating that the small dipole is totally reactive, that is to say, it does not radiate any field. In the presence of the resonant metalens, the curve decreases down to a value of 0.7, meaning that the source is better matched due to the Purcell effect as stated before. The better coupling between the source and the resonant medium permits to compensate the less efficient radiation of the most subwavelength modes: the two effects counterbalance themselves and each mode eventually radiate the same amount of energy.

\section{Conclusion}

In this article, we have explained the physical mechanism of our original paper concerning the concept of resonant metalens \cite{metalens}. The intuitive property of resonance splitting when considering coupled resonators has been investigated through a wave approach. Starting from the wavefield inside an infinite wire medium, we have seen that a wire medium with finite dimensions can be seen as a macro resonator that contains many resonance frequencies.  Interestingly, each of these resonances corresponds to an eigen wavefield that fluctuates on its own sub-wavelength scale. Then, we have shown how these modes can radiate into the far field region information that is usually contained in the evanescent spectrum. Eventually, by placing such a resonant medium near a sub-wavelength source we have seen that the Purcell effect permits an efficient radiation. 

These physical mechanisms permit to get information of a source on a smaller scale than the wavelength, stored as a time/frequency signature, in the far field region. In the next article we will explain how one can use these properties for breaking the diffraction barrier in a focusing/imaging scheme and study the main limitations as well as losses impact.  

\appendix*
\section{Sophisticated dispersion law}

When calculating the dispersion law of guided modes inside the wire medium, we have took sides for neglecting both the TM evanescent waves and the high orders of diffraction generated at the reflection. For the range of frequency considered, the TM waves are strongly evanescent because the plasma wave number $k_p$ is really higher than any other wave number considered \cite{belovTM,belov2}. And the high order of diffractions are also strongly evanescent because the fundamental one is already evanescent. All of these waves should modify the reflection coefficient of equation (\ref{r}) by an additional factor which affects phase, and thus dispersion. 

A dispersion law of the guided modes considering the TM waves is presented in the appendix of ref. \cite{PREBelov}. To obtain this dispersion law the authors used an effective medium approach which leads to an effective anisotropic permittivity along the wire axis\cite{belov2}. It permits to obtain the Bloch wave number of the TM modes: $\kappa_\textrm{TM}=\sqrt{k^2-k_\perp{}^2-k_p{}^2}$, which is a complex quantity for the range of frequency considered here. To keep the formalism of their article, it is  convenient  to  use  the  notation  $\kappa_\textrm{TM}=i \gamma_\textrm{TM}$. Also,  by  analogy,  we  denote  the quantity $\kappa=\sqrt{k^2-k_\perp{}^2}=i \gamma_0$, where $\gamma_0$ is a real quantity for the waves we consider here (evanescent waves in free space). Then, an Additional Boundary Condition is introduced to solve the boundary problem at the air/wire medium interface \cite{silveirinha}, and the law obtained can be summarized in:

\begin{equation}\label{dispersion_complex}
\tan\left(k\frac{L}{2}\right)=\frac{\gamma_\textrm{app}}{k}
\end{equation}

\noindent with,

\begin{equation}
\gamma_\textrm{app} = \gamma_0\frac{k_\perp{}^2+k_p{}^2}{k_p{}^2} +\gamma_\textrm{TM}\frac{k_\perp{}^2}{k_p{}^2}\tanh(\frac{\gamma_\textrm{TM}L}{2}) 
\end{equation}

\begin{figure}
\includegraphics[width=8.5cm]{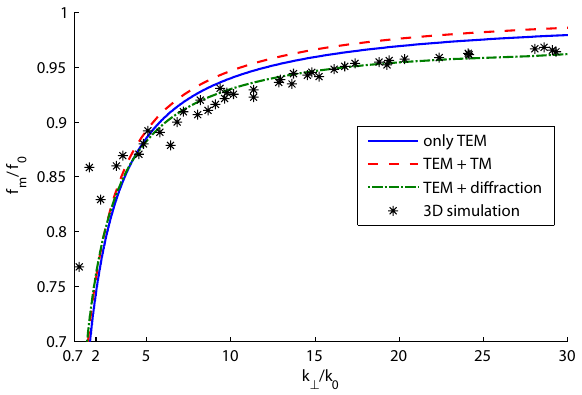}
\caption{\label{dispersion_with_TM} The dispersion laws of the first resonance frequencies $f_m$ (normalized to the original Fabry-Perot frequency $f_0$) versus the transverse Bloch wave number $k_\perp$ (normalized to the original Fabry-Perot wave number $k_0$) from equation (\ref{dispersion1}) only considering TEM waves in the wire medium (blue solid line), from ref. \cite{PREBelov} with our parameters considering TM and TEM waves (red dashed line), and from equations (\ref{dispersion_complex}) and (\ref{diffraction}) considering only TEM waves but high diffraction orders (green dashed dotted line). Superimposed with stars is the results obtained from 3D simulations. }
\end{figure}

This dispersion relation is actually the same as the one presented in the manuscript with a correction due to the TM waves which appears in $\gamma_\textrm{app}$. In the limit of thin wires and for frequencies well below the cutoff of TM waves (which is the case considered here), the plasma wave number $k_p$ is extremely higher than any other wave number considered. In that limit case we find that $\gamma_\textrm{app}=\gamma_0$ and the dispersion law presented in equation (\ref{dispersion1}) remains true. With the geometrical parameters we took in this example, the difference between the two laws is less than 1 \%   (Fig. \ref{dispersion_with_TM}): this is the reason why we chose not to take into account the strongly evanescent TM waves. 

As seen in the expression of $\gamma_\textrm{app}$ which is always greater than $\gamma_0$, the TM waves tend to increase the apparent wave number associated to the evanescent decrease in air, or equivalently to decrease the penetration depth inside air. The guided modes are much more weakly confined to the interface than the case without considering TM waves. As a consequence, for a given transverse wavenumber $k_\perp$ it results in a higher frequency for the guided mode than the dispersion law presented in the paper.

We would like to argue that the dispersion law obtained with 3D time domain simulations (Fig. \ref{dispersion_with_TM}) gives a distinct comportment for large $k_\perp$ than the one expected when adding the TM waves. Indeed, while the correction due to the TM waves tend to increase the frequency, the simulation results show that a decrease is expected. Actually, the periodic nature of the wavefield and thus the Bragg waves have been neglected, a remaining problem with effective medium approach. Here we propose to do the calculation of the dispersion law taking into account some high orders of diffraction with a method inspired from ref. \cite{Vidal,Hibbins_calcul}. 

We consider a TEM Bloch wave with transverse wave number $\mathbf{k_\perp}=k_\perp\mathbf{e_x}$ traveling in the half space $z<0$ filled out with a wire medium. This Bloch wave encounters a boundary with vacuum at $z=0$ and gives rise to a reflected TEM Bloch wave inside the wire medium region. The superposition of the 2 TEM Bloch waves writes at $z=0$:

\begin{eqnarray}
\mathbf{E}_\textrm{TEM}&=&(1+r)e^{i k_\perp x} \left( \begin{array}{c} F(x,y)\\ 0\\ 0\\ \end{array} \right) \nonumber \\
\mathbf{H}_\textrm{TEM}&=&\frac{1}{\mu_0 c} (1-r) e^{i k_\perp x}\left( \begin{array}{c} 0\\F(x,y)\\ 0\\ \end{array} \right) 
\end{eqnarray}

The function $F(x,y)$ stands for the Bloch envelope of the TEM wave inside the wire medium, and notably this function is null over the wire section of a unit cell. One can notice that the $y$ (respectively $x$) component of the electric (respectively magnetic) component of the TEM wavefield has been neglected: the diffraction orders along the $y$-axis are neglected. This implies that the reflected and transmitted TE modes will be neglected too, but in the general case they must be taken into account. We will also neglect the TM reflected waves because their influence has been discussed earlier. Thus, the transmitted field in the vacuum region, with these approximations, writes at $z=0$:

\begin{eqnarray}
\mathbf{E}_\textrm{vac}&=&\sum_{n=-\infty}^{\infty}t^{(n)} e^{i k_\perp^{(n)} x} \left( \begin{array}{c} k_z^{(n)}/k \\ 0\\ -k_\perp^{(n)}/k\\ \end{array} \right) \nonumber \\
\mathbf{H}_\textrm{vac}&=&\frac{1}{c} \sum_{n=-\infty}^{\infty}\frac{t^{(n)}}{\mu_0 c} e^{i k_\perp^{(n)} x}\left( \begin{array}{c} 0\\1\\ 0\\ \end{array} \right) 
\end{eqnarray}

\noindent with $k_\perp^{(n)}=k_\perp+\frac{2 n \pi}{a}$ and in the range of frequency considered, we have again $k_z^{(n)}=i\gamma_n=i\sqrt{{k_\perp^{(n)}}^2-k^2}$. 

To eliminate the unknowns, we use the fact that the tangential components of the electric field  must  be  continuous  at  the  vacuum/wire medium interface over the entire unit cell (which represents the area $\Omega$), while the  magnetic field  components  are  continuous  only  at  the air regions of the unit cell (the area $\Omega$ minus the area $\mathcal{D}$ corresponding to the wire section). Matching the $x$ component of the electric field, multiplying by $e^{-i k_\perp^{(m)} x}$, and integrating over the area $\Omega$ yields:

\begin{equation}
t^{(m)}=\frac{k}{k_z{}^{(m)}}(1+r) I_m
\end{equation}

\noindent where,

\begin{equation}
I_m=\frac{1}{a^2}\iint_\Omega F(x,y) e^{i k_\perp x} e^{-i k_\perp^{(m)} x} \textrm{d}S
\end{equation}

Again, matching the $y$ component of the magnetic field, multiplying by $F(x,y) e^{-i k_\perp x}$, and integrating over the area $\Omega-\mathcal{D}$ gives:

\begin{equation}
\sum_{n=-\infty}^{\infty} t^{(n)} I_n{}^*= (1-r) J_0
\end{equation}

\noindent where,

\begin{equation}
J_0=\frac{1}{a^2}\iint_{\Omega-\mathcal{D}} F(x,y)^2 \textrm{d}S
\end{equation}

\noindent Then we can extract the reflection coefficient of the TEM wave, and as done in the manuscript when adding a second along the wires, it eventually gives the same dispersion law as equation (\ref{dispersion_complex}) with in this case:

\begin{equation} \label{diffraction}
\gamma_\textrm{app}=\frac{J_0}{\sum\limits_{n=-\infty}^{\infty} \frac{|I_n|^2}{\gamma_{n}{}^2}}
\end{equation}

To evaluate this quantity, we need to give an expression for the function $F(x,y)$. For simplicity we have chosen that this function is equal to $1$ in the domain $\omega-\mathcal{D}$, and null in the domain $D$. The dispersion law obtained considering the 2 first diffraction orders ($+1$ and $-1$) is presented in figure \ref{dispersion_with_TM}. One can notice that this dispersion law gives a better fit to the results obtained from the full 3D simulations. Obviously, this calculation is not an exact one but it has the merit to show that taking into account the diffraction orders tends to increase the apparent penetration depth inside air, thus giving a better comportment for the high transverse wave numbers. Nevertheless, this calculation greatly complexifies the understanding for giving a correction less than 1 \%: this is reason why we also neglect the high diffraction orders during the whole manuscript.

\bibliography{biblioPRB}

\end{document}